\documentclass[12pt]{iopart}

\usepackage{graphicx}

\newcommand{\beq}{\begin{equation}}
\newcommand{\eeq}{\end{equation}}
\begin{document}

\title{Flux-flow and vortex-glass phase in iron pnictide BaFe$_{2-x}$Ni$_x$As$_2$ single crystals with $T_c$ $\sim$ 20 K}
\author{S Salem-Sugui Jr.$^1$, A D Alvarenga$^2$, H-Q Luo$^3$, R Zhang$^3$ and D-L Gong$^3$}

\address{$^1$ Instituto de Fisica, Universidade Federal do Rio de Janeiro, 21941-972 Rio de Janeiro, RJ, Brazil}

\address{$^2$ Instituto Nacional de Metrologia Qualidade e Tecnologia, 25250-020 Duque de Caxias, RJ, Brazil}

\address{$^3$ Beijing National Laboratory for Condensed Matter Physics, Institute of Physics, Chinese Academy of Sciences, Beijing 100190, China}

\ead{said@if.ufrj.br}

\begin{abstract}
We analysed the flux-flow region of isofield magneto resistivity data obtained on three crystals of BaFe$_{2-x}$Ni$_x$As$_2$  with $T_c$$\sim$20 K for three different geometries relative to the angle formed between the applied magnetic field and the c-axis of the crystals.  The field dependent activation energy, $U_0$, was obtained from the TAFF and modified vortex-glass models, which were compared with the values of $U_0$ obtained from flux-creep available in the literature. We observed that the $U_0$ obtained from the TAFF model show deviations among the different crystals, while the correspondent glass lines obtained from the vortex glass model are virtually coincident. It is shown that the data is well explained by the modified vortex glass model, allowing to extract values of $T_g$, the glass transition temperature, and $T^*$, a temperature which scales with the mean field critical temperature $T_c(H)$. The resulting glass lines obey the anisotropic Ginzburg-Landau theory and are well fitted by a theory developed in the literature by considering the effect of disorder. 

\end{abstract}

\submitto{\SUST}
\maketitle

\section{Introduction}
The study of the flux-flow region in type II superconductors gained additional attention after the discovery of the high-$T_c$ superconductors \cite{bednorz} which display a large reversible region associated to a non-zero resistivity. This non-zero resistivity flux-flow region has been mainly explained in terms of the thermal assisted flux-flow model \cite{kess}, TAFF, within a vortex-liquid phase that forms as the vortex lattice is thermally depinned as temperature increases above the irreversibility line. Moreover, it has been shown that depending on the disorder, a vortex-glass phase, VG, can emerge below the glass temperature, $T_g$, at which the resistivity is zero \cite{baruch1,fisher}. A finger-print of this phase is that the flux-flow resistivity should exhibit a critical behaviour as temperature approaches $T_g$. As a consequence of these two models, the flux-flow region usually studied from isofield magneto-resistivity curves, allows to obtain from the theories the activation energy, $U_0$($H$), associated with the dissipative mechanism which is an important parameter from the potential applications point of view.  As iron pnictides systems \cite{japan} also display a considerable large reversible region, the flux-flow region have been studied in many compounds, including Ca$_{0.82}$La$_{0.18}$FeAs$_2$ \cite{wei1}, CeFeAsOF \cite{1,2}, SmFeAsO$_{0.9}$F$_{0.1}$ \cite{3}, BaFe$_{2-x}$Ni$_x$As$_2$  \cite{shahbazi,4}, BaFe$_{2-x}$Co$_x$As$_2$ \cite{arx}, (Ba,K)Fe$_2$As$_2$ \cite{arx2}, the chalcogenades FeTe$_{0.60}$Se$_{0.40}$ \cite{5} and Fe$_{1.04}$Te$_{0.6}$Se$_{0.4}$ \cite{5b}, and Ag-doped FeSe$_{0.94}$ \cite{6} among others. On the other hand, studies of the vortex-glass phase in pnictides are still reduced being restricted to few compounds.\cite{ghorbani,shahbazi,prando,hao}  It should be noted that most of the flux-pinning studies in pnictides are conducted in only one sample of each system. Since pnictide superconductors has been considered as a potential materials for applications, a question that rises  is whether flux-pinning properties as well as the irreversibility line show changes among different samples of the same system. 
 
In the present work, we address the above mentioned important issue by studying the flux-flow region on three crystals of the electron-doped pnictide system of BaFe$_{2-x}$Ni$_x$As$_2$ \cite{first}, which is one of the most studied pnictide systems \cite{shahbazi,4,7,8,9,luo,10} and may be considered a potential material for application in future. The study was conducted by analysing isofield magnetoresistivity data obtained for three different geometries of the applied magnetic field with respect to the c-axis of the samples, on three different high-quality crystals with superconducting transitions temperature, $T_c$, equal to 19.8 K (crystal$\#$1), 19.7 K (crystal $\#$2) and 19.8 K (crystal $\#$3), and $x$ equal to 0.096 ($\#$1), 0.096($\#$2) and 0.098 ($\#$3) respectively, where the values of $x$ are  close to the optimally doped $x$=0.1. In parallel with this comparative study of flux-flow region in three different samples of same pnictide system, we have also made some efforts to see the possibility of vortex-glass phase. Another motivation was try to compare values of $U_0$ obtained in the flux-flow region with those obtained in the irreversible region from flux-creep measurements. Such comparison can be useful as many samples are characterised from flux-creep measurements. The flux-flow region of each isofield curve was analysed in terms of the TAFF \cite{kess} and the modified vortex-glass models \cite{rapp1,rapp2}, allowing to obtain the correspondent values of the field dependent activation energies. Further, these values of the activation energy were compared with the $U_0$ extracted from the flux-creep data presented in the literature for BaFe$_{2-x}$Ni$_x$As$_2$ superconductor with $T_c$$\sim$20 K \cite{sust1}. We observed that $U_0$($H$) obtained from flux-creep are about an order of magnitude smaller than those obtained from the TAFF model, but are of same order of magnitude of those obtained from the modified vortex-glass model. It has also been observed that values of $U_0$($H$) obtained from the TAFF model for the present three crystals show small deviations among them, while plots of the glass lines obtained for these three crystals are virtually identical. We observed that our data obey the modified vortex-glass model allowing to estimate values of $T_c$($H$), which is a new result. It is shown that the resulting glass lines obey the Ginzburg-Landau anisotropic theory and are well fitted by a theory developed in ref.\cite{baruch2} for the glass line considering the effect of disorder.
\section{experimental}
The temperature dependence of the resistivity, $\rho$($T$) were obtained by the standard four probe ac lock-in technique with the current flowing along the ab-planes (along the b-axis) of the crystals always perpendicular to the applied magnetic field, and voltage was measured with wires placed along the a-axis. Isofield magneto-resistivity curves were obtained as a function of temperature for applied magnetic fields, $H$, running from 0 to 14 T, for three different orientations of $H$ with respect to the c-axis of the samples: $H$$\parallel$c-axis, $H$$\perp$c-axis ($H$$\parallel$ab-planes) and $H$ forming an angle of 53 degrees with the c-axis. Information regarding samples grown can be found in ref.\cite{luo} and details about samples dimensions and resistivity measurements can be found in ref.\cite{jesus1} where the same data here studied were analysed in terms of superconducting fluctuations in the conductivity. It is worth mentioning that the analysis presented in ref.\cite{jesus1} was focused on data lying above the superconducting transition temperature with field, $T_c(H)$, corresponding to data lying above the flux-flow region.

\section{results and discussion}
Figure 1 shows selected logarithm  plots of the magneto-resistivity curves against 1/$T$, (arrhenius plots) as obtained for the three studied crystals for the three different field orientations with respect to the c-axis of the crystals. Schematics of the geometric configuration of the resistivity measurements are shown in each figure.  In all panels of Fig. 1 the lower part of each curve shows a well defined linear behaviour that has been interpreted in terms of  the well know TAFF model.\cite{kess} According to the TAFF model, the activation energy $U_0$ may be estimated using the relation,  ln$\rho$ = ln$\rho_0$-$U_0$/$k_B$$T$, where $k_B$ is the Boltzmann constant. The $T_c$ for each crystal is defined as the temperature for maximum in the $d\rho/dT$ at zero field (inset of Fig. 1a) where $\Delta T_c\leq0.3$~K for each crystal, evidencing the sharp superconducting transition of the crystals. The insets of Fig.1b and Fig.1c show selected plots of [d(ln$\rho$)/d$T$]$^{-1}$ vs. $T$ for crystals $\#$2 and $\#$3 data. The linear behaviour observed in the low temperature side of the $[dln\rho/dT]^{-1}$ vs. $T$ curve, may be defined in terms of the vortex-glass model. According to the vortex-glass model, the resistivity near the vortex-glass state may be defined in terms of the power law,
\begin{equation}
\rho = \rho _n \mid (T-T_g)/T_g \mid ^s ,
\end{equation}
where $T_g$ is the glass transition temperature, $\rho _n$ is related to the normal resistivity and $s$ is the glass critical exponent. The linear behaviour of the curves shown in the inset  of Fig. 1(b) and Fig. 1(c) was observed for all $\rho$ vs. $T$ curves of this work, allowing to obtain the vortex-glass line for each crystal for the three geometries of the experiment. The upper limit of the VG linear region ends at the temperature $T_{cros}$, which marks the crossover from the VG linear region to the vortex liquid phase, above which the TAFF model applied. We mention that the plots of [d(ln$\rho$)/d$T$]$^{-1}$ vs. $T$ yield values of $s$ varying from 2.5 to 3.1, that is in reasonable agreement with the expected value for the vortex-glass transition, $s$ = 2.8 \cite{fisher}.

\begin{figure}[t]
\begin{center}
\includegraphics[scale=.4]{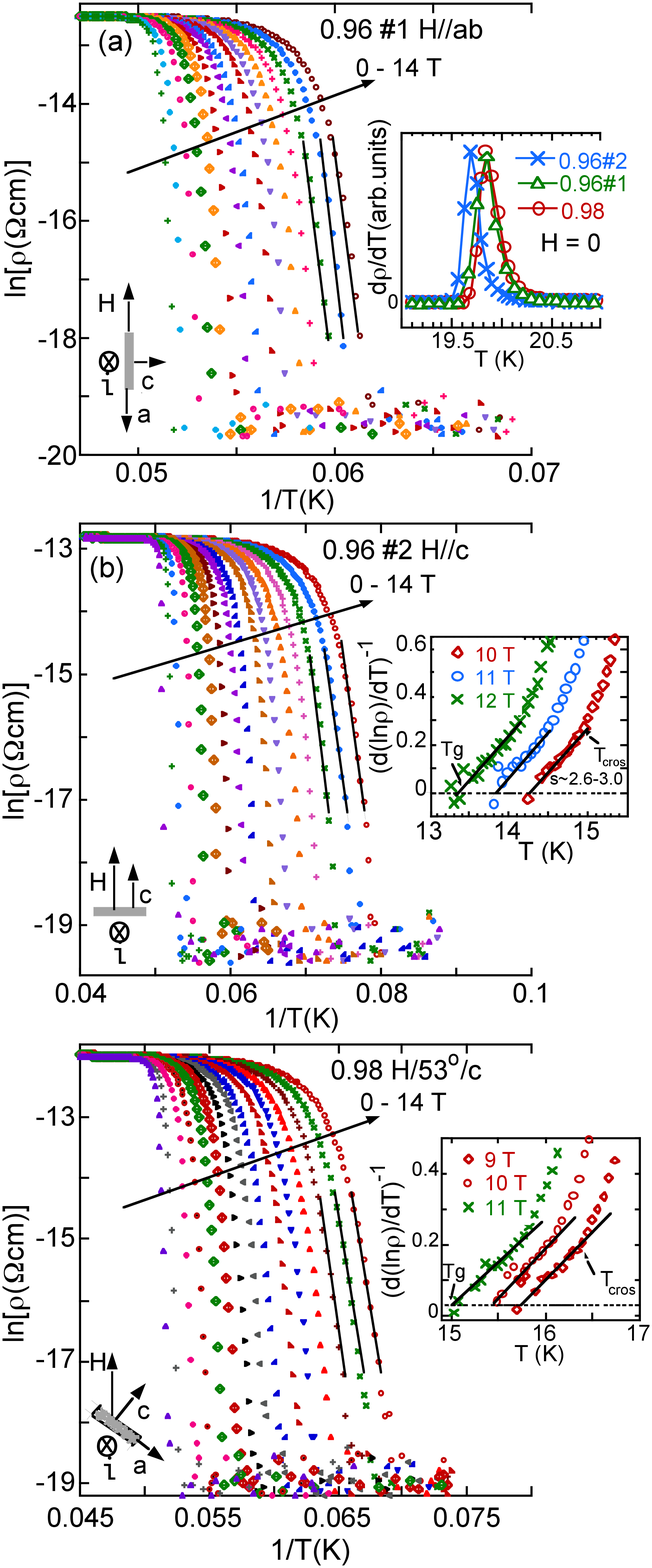}
\caption{selected plots of ln($\rho$vs.1/$T$ for fixed fields: a) crystal $\#$1 for $H$$\parallel$c-axis; b) crystal $\#$2 for $H$$\parallel$ab-planes; and c) crystal $\#$3 for $H$ forming 53$^o$ with the c-axis. Schematics of the geometric configuration of the resistivity measurement are shown inside each figure.  Insets: a) detail of the superconducting transition of the three crystals; b) and c) selected plots of  [d(ln$\rho$)/d$T$]$^{-1}$ vs. $T$ for data of the correspondent main figures. Solid lines in the main figures and in the insets are only a guide to the eyes.}
\label{fig1}
\end{center}
\end{figure}

It is interesting to compare the values of $U_0$($H$) extracted from Fig. 1 using the TAFF model, with those obtained from flux-creep measurements in ref.\cite{sust1} for a BaFe$_{2-x}$Ni$_x$As$_2$  sample with similar content of Ni and approximately same $T_c$$\sim$20 K. In this case, values of $U_0$ = $k_B$$T$/$R$, where $R$ is the relaxation rate, can be directly estimated from values of $R$ = (1/$M_0)$d$M$/dlnt $\sim$ d(ln$M$)/dlnt presented in  Ref.\cite{sust1} for $H$$\parallel$ c-axis and for $H$$\parallel$ ab-planes. In order to make a more complete comparison between values of $U_0$ obtained from different approaches, we have also applied the modified vortex-glass model \cite{rapp1,rapp2,shahbazi} to the $\rho(T)$ data of crystal $\#$2 (sample with the sharpest $T_c$). The modified vortex-glass model  considers the differences in temperature $T$-$T_g$ near the glass transition to be directed related to the difference in energy $k_B$$T$-$U_0$, where $U_0$ is an effective activation energy which is a function of $T$ and $H$. Then, by replacing $T$/$T_g$ by $k_B$$T$/$U_0$ in Eq. 1 it is possible to obtain the following equation for $U_0$ 
\begin{equation}
 U_0(H,T)= k_B T[1+(\rho/\rho _n)^{1/s}]^{-1}
\end{equation}
Figure 2 shows isofield curves of $U_0$ vs. $T$ as calculated using Eq. 2 for $\rho(T)$ data of crystal $\#$2 data for $H$$\parallel$c-axis (Fig.2a) and for  $H$$\parallel$ab-planes (Fig.2b), using $s$ = 2.8. As shown in Fig.2, the lower part of the many different curves shows a linear behaviour which encounters the upper inclined line formed by the collapse of the many curves at $T_g$  (as shown by the arrows for $H$ = 14 $T$) and encounters the x-axis at a temperature $T^*$($H$). It should be noted that the same linear behaviour as in Fig. 2a was observed for YBa$_2$Cu$_3$O$_{7- \delta}$ \cite{rapp1,rapp2} and for BaFe$_{1.9}$Ni$_{0.1}$As$_2$ with $T_c$=18.3 K ref.\cite{shahbazi} but with all extended lines encountering the x-axis at $T_c$. The linear behaviours shown in Figs. 2a and 2b which extends to  $T^*$ suggest an empiric expression for the effective pinning energy given by 
$U_0$($H,T$) = $U_b$(1-$T$/$T^*$), where $U_b$ is the field dependent pinning energy at $T$ = 0. Regarding the values of $T^*$, we observed that these values scale with values of $T_c(H)$ extracted from the respective resistivity curves assuming that the transition occurs for $\rho$ corresponding to 80 $\%$ of the normal resistivity. The inset of Fig.2b shows a plot of $T^*(H)$ along with the correspondent values of $T_c(H)$ for data of Fig.2, evidencing the same behaviour of these two quantities ($T^*$ and $T_c(H)$) for both field directions. The curves shown in the inset of Fig. 2b yield d$H_{c2}$/d$T$ = -2.98 T/K for $H$$\parallel$c-axis and -7.47 T/K for $H$$\parallel$ab-planes which by using the WHH formula \cite{whh}, $H_{c2}(0)$=-0.693$T_c$d$H_{c2}$/d$T$, yield: $H_{c2}(0)$ = 41 T for $H$$\parallel$c-axis and 102 T for $H$$\parallel$ab-planes, corresponding to  an anisotropic factor $\gamma$ = 2.5. Values of d$H_{c2}$/d$T$ found above for both field directions are in strict agreement with values listed in ref.\cite{luo2} for a BaFe$_{2-x}$Ni$_x$As$_2$ crystal with $x$=0.096 and $T_c$=19.9 K.

\begin{figure}[t]
\begin{center}
\includegraphics[scale=.4]{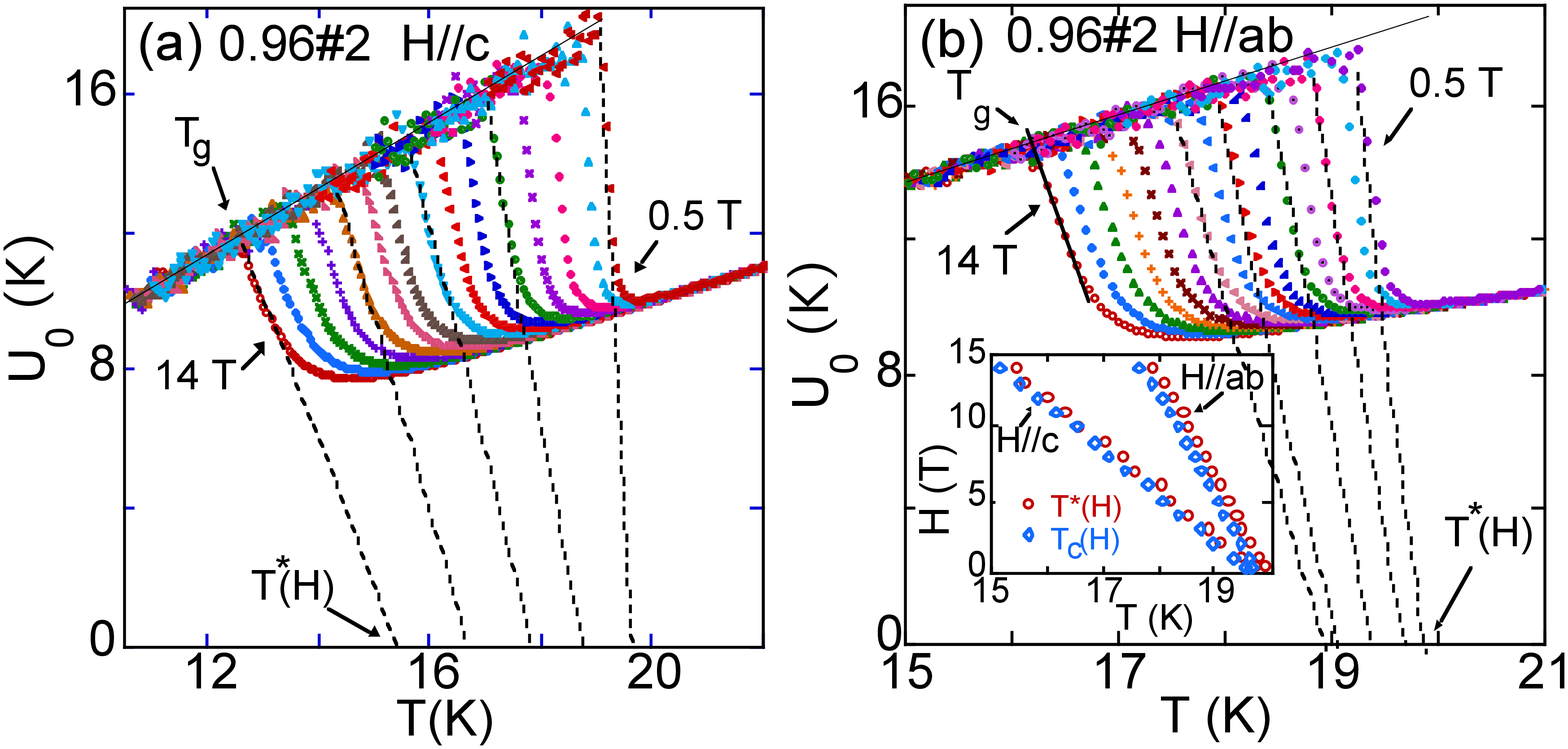}
\caption{Plots of $U_0$/$k_B$ vs. $T$ obtained for crystal $\#$2 data by using Eq. 2: a) $H$$\parallel$c-axis; b) $H$$\parallel$ab-planes. The inset of Fig. 2b shows a plot of $T^*$($H$) along with values of $T_c(H)$ for both field directions}
\label{fig2}
\end{center}
\end{figure}

With the empiric expression $U_0$($H,T$) = $U_b$(1-$T$/$T^*$) obtained from Fig. 2, and substituting  $U_0$=k$_B$$T_g$ at $T$=$T_g$, one can obtain $U_b$=k$_B$$T_g$$T^*$/($T^*$-$T_g$), producing $U_0$=k$_B$$T_g$($T^*$-$T$)/($T^*$-$T_g$) which can be substitute in Eq. 2 resulting:
\begin{equation}
 \rho /\rho _n=[(T/T_g)(T^*-T_g)/(T^*-T)-1]^s.
\end{equation} 
The above equation is expected to be an universal form for the resistivity within the vortex-glass to vortex-liquid phase, where  [($T$/$T_g$)($T^*$-$T_g$)/($T^*$-$T$)-1] = $t_s$ is the scaled temperature. Then double-logarithmic plots of $\rho$/$\rho _n$ vs. $t_s$ obtained for different fields are expected to collapse on a single curve. Figure 3 shows these plots as obtained for the same data presented in Fig.2 for $H$$\parallel$c-axis (Fig.3a) and for $H$$\parallel$ab-planes (Fig.3b) evidencing a collapse of all different curves. It is important to note that the quantity in the x-axis of Fig.3 differs from the similar scaled temperature $t_s$  presented in refs.\cite{rapp1,rapp2,shahbazi} ,  [($T$/$T_g$)($T_c$-$T_g$)/($T_g$-$T$)-1], which was obtained by assuming the empiric expression of $U_0$=$U_b$(1-$T$/$T_c$). To show that the later $t_s$ used in refs.\cite{rapp1,rapp2,shahbazi} for plots of $\rho$/$\rho _n$ is not appropriate for our data, we plot in the insets of Fig.3, $\rho$/$\rho _n$ vs. [($T$/$T_g$)($T_c$-$T_g$)/($T_g$-$T$)-1] as obtained for the same data presented in each main figure. It is possible to see in these insets, a poor collapse of the curves for $H$$\parallel$c-axis, and no collapse at all for $H$$\parallel$ab-planes. Another consequence of the obtained empiric expression of $U_0$ is that it is not possible to obtain a functional expression for the glass line, $H_g$($T$), as the obtained expression after follow the same procedure presented in ref\cite{rapp1} and ref\cite{rapp2} is dependent of $T^*$.

\begin{figure}[t]
\begin{center}
\includegraphics[scale=.5]{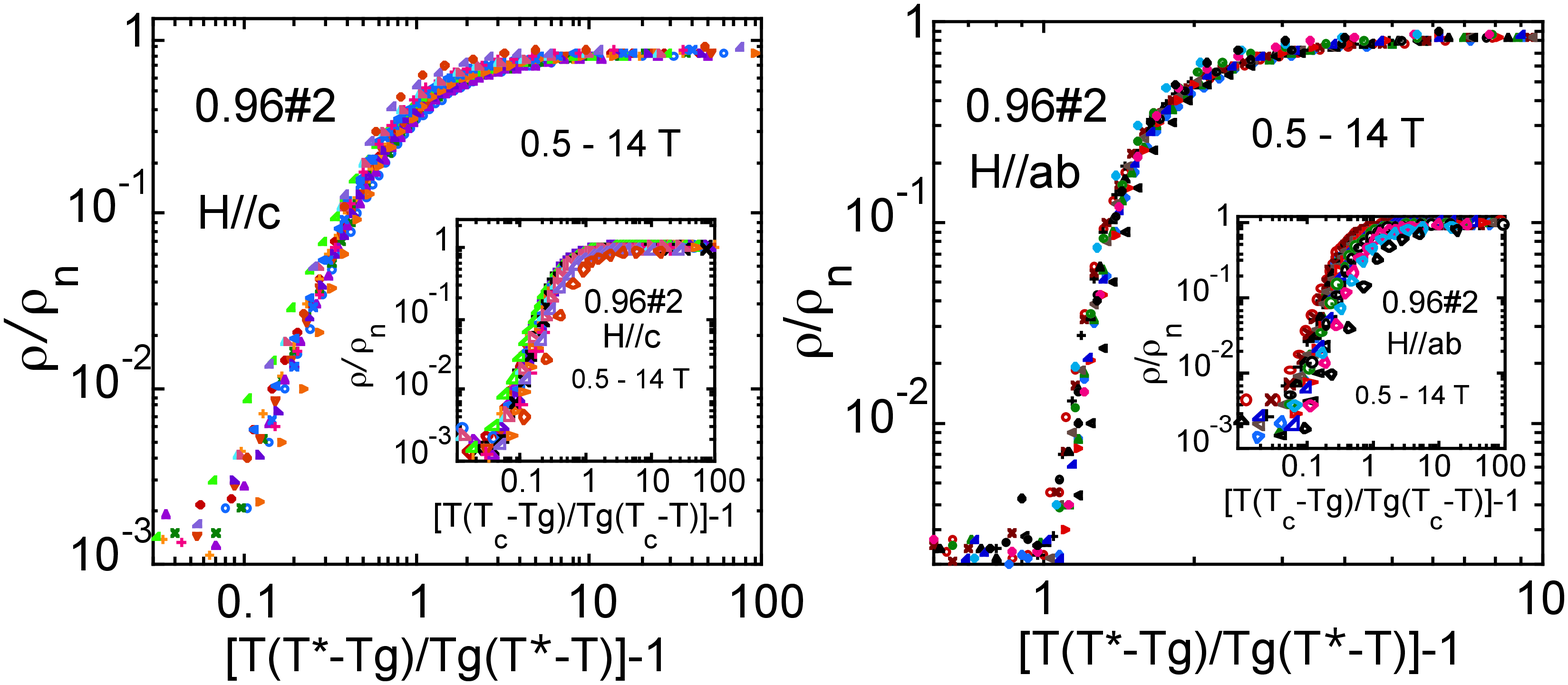}
\caption{Plots of $\rho$/$\rho _n$vs. $T_s$ for crystal $\#$2 data, where $T_s$ was obtained from the empiric form of $U_0$($H,T$) = $U_b$(1-$T$/$T^*$) extracted from Fig.2: a) $H$$\parallel$c-axis; b) $H$$\parallel$ab-planes. The insets show the same plots of the main figures but for a $T_s$ obtained by assuming  the empiric form of $U_0$=$U_b$(1-$T$/$T_c$)}
\label{fig3}
\end{center}
\end{figure}

We plot in Fig.4 the values of $U_0$ as obtained from the TAFF model for the three crystals along with values of $U_b$ obtained from a linear fitting conducted on data lying in the linear regions shown in Fig.2 for crystal $\#$2. We also plot in Fig. 4 values of $U_0$ obtained from flux-creep in a BaFe$_{2-x}$Ni$_x$As$_2$ crystal with similar $T_c$ for $H$$\parallel$c-axis at $T$ = 8 K and for $H$$\parallel$ab-planes at T = 13 K. \cite{sust1} Figure 4 evidences that the behaviour of $U_0$ obtained from TAFF model is almost sample independent for $H$$\parallel$c-axis (Fig.4a), showing a small deviation for crystal $\#$1 for $H$$\parallel$ab-planes (Fig.4b). As usually observed \cite{shahbazi,wei1,5b}, the double-logarithmic plots of Fig.4 show two distinct linear behaviours for $U_0$ obtained from the TAFF model and also for $U_b$ obtained from the modified vortex-glass model, where the field $H$$\sim$ 4 T separates the low field region (showing a lower dependence with field) from the high field region (showing a higher dependence with field). As discussed in Ref.\cite{wei1}, it is believed that the double linear field dependence shown in Fig. 4 is associated with a transition from single-vortex pinning in the low field region to a small bundle pinning in the high field region. As observed in Ref.\cite{shahbazi}, values of $U_b$ in Fig. 4 follow the same trend as observed for the $U_0$ values obtained from the TAFF model but with values of about one order of magnitude smaller. We observed that the power law exponents associated with the linear behaviours are higher for $H$$\parallel$c-axis which for instance for crystal $\#$2, 
$U_0$$\sim$$H$$^{-0.41}$ and $U_b$$\sim$$H$$^{-0.48}$ for the low field region below 4 T and  $U_0$$\sim$$H$$^{-1.2}$ and $U_b$$\sim$$H$$^{-0.93}$ for the high field region while for $H$$\parallel$ab-planes, $U_0$$\sim$$H$$^{-0.16}$ and $U_b$$\sim$$H$$^{-0.25}$ for the low field region and  $U_0$$\sim$$H$$^{-0.94}$  and $U_b$$\sim$$H$$^{-0.68}$ for the high field region. This small anisotropy of $U_0$, showing lower exponents of the power law behaviour for $H$$\parallel$ab-planes has also been observed for Fe$_{1.04}$Te$_{0.6}$Se$_{0.4}$ \cite{5b} and for Ca$_{0.82}$La$_{0.18}$FeAs$_2$ \cite{wei1}. It is interesting to observe that the values of $U_0$ obtained in the irreversible regime, from flux-creep, are about one order of magnitude smaller than those obtained in the reversible regime from the TAFF model, but most important, they are about the same magnitude of $U_b$ obtained from the modified vortex-glass model. The later suggests that the modified vortex-glass model can capture the activation energy experienced in the flux-creep regime. It is possible to see that the values of $U_0$ obtained from flux-creep presents a maximum, which is absent in the $U_0$ obtained using TAFF model as well in $U_b$. These maximums in $U_0$ obtained from flux-creep are related to minimums observed in the respective plots of $R$ vs. $H$ presented in ref.\cite{sust1}. Despite the minimums positions in that case are not direct related to neither the position of $H_p$ (the second magnetisation peak position, SMP) nor $H_{on}$ (the onset of the SMP) in isothermic $M(H)$ curves, they are related, as discussed in ref.\cite{wei2}, to a crossover from elastic to plastic pinning occurring as field increases above the minimum in $R$ (the maximum in $U_0$), which mechanism explain how the SMP develops in $M$($H$) curves. The upper inset of Fig. 4b shows a general non-linear form attributed to the isofield activation energy $U$($J$) as a function of the superconducting current, where a tangent to the curve for a given current $J1$ intercept the y-axis at $U_0$($J1$) which represents the $U_0$ activation energy determined from flux-creep measurements \cite{maley}. As shown in this inset, as temperature increases $J$ decreases and the value of $U_0$ increases. As a consequence, as $J$ approaches zero with $T$ approaching $T_{irr}$, $U_0$ is expect to reach a much larger value than values obtained in flux-creep experiments. One would expect that the value of $U_0$($J$$\rightarrow$0) should be of the same order of $U_0$ obtained from the TAFF model in the flux flow region where $J$$=$0. This conjecture may explain why values of $U_0$ obtained from the TAFF model are much larger than $U_0$ values obtained from flux-creep. The inset of Fig.4a exemplify the behaviour of $U_0$ (TAFF) with respect to the field orientation as observed for each studied samples, where it is possible to see that  only the direction $H$$\parallel$ab-planes shows a considerably change in the values of $U_0$. The lower inset of Fig.4b shows a plot of $U_0$ (from TAFF model) as obtained for the three crystals when $H$ formed an angle of 53 degrees with the c-axis, evidencing small deviations in the values of $U_0$ among the different crystals for this geometry.

\begin{figure}[t]
\begin{center}
\includegraphics[scale=.4]{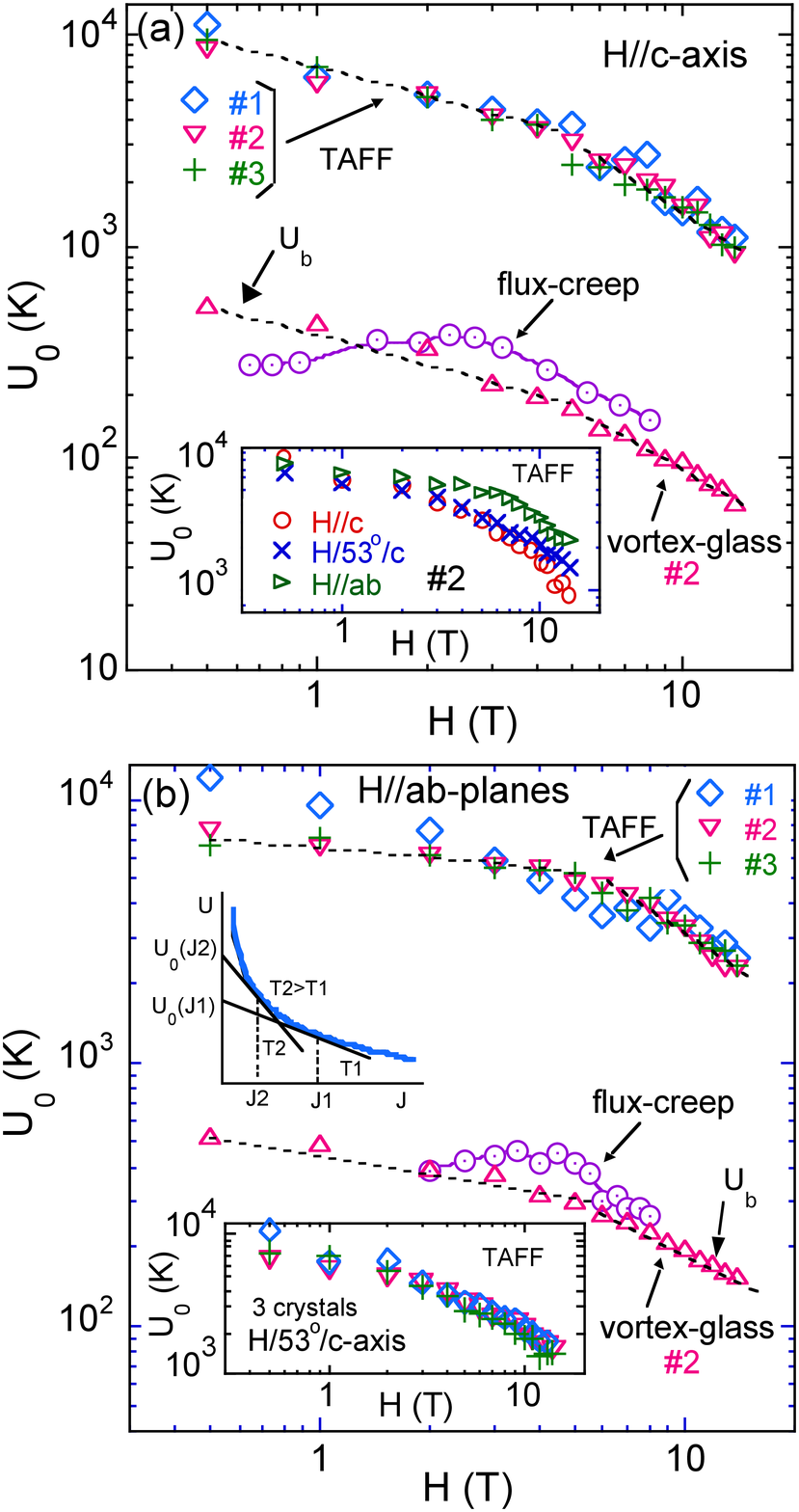}
\caption{Plots $U_0$ vs. $H$ obtained from TAFF model for the three crystals, along with $U_0$ vs. $H$ extracted from flux-creep and $U_b$ vs. $H$ for crystal $\#$2 obtained from the vortex glass model: a) $H$$\parallel$c-axis; b) $H$$\parallel$ab-planes. Insets: a) plots of $U_0$ vs $H$ from TAFF model for crystal $\#$2 for the three different geometries; b) upper: non linear form of $U$ vs $J$ for a fixed field; lower: plots of $U_0$ vs $H$ from TAFF model for $H$ forming an angle of 53$^o$ with the c-axis for the three crystals. Dotted and solid lines appearing in the main figures are only a guide to the eyes.}
\label{fig4}
\end{center}
\end{figure}

\begin{figure}[t]
\begin{center}
\includegraphics[scale=.4]{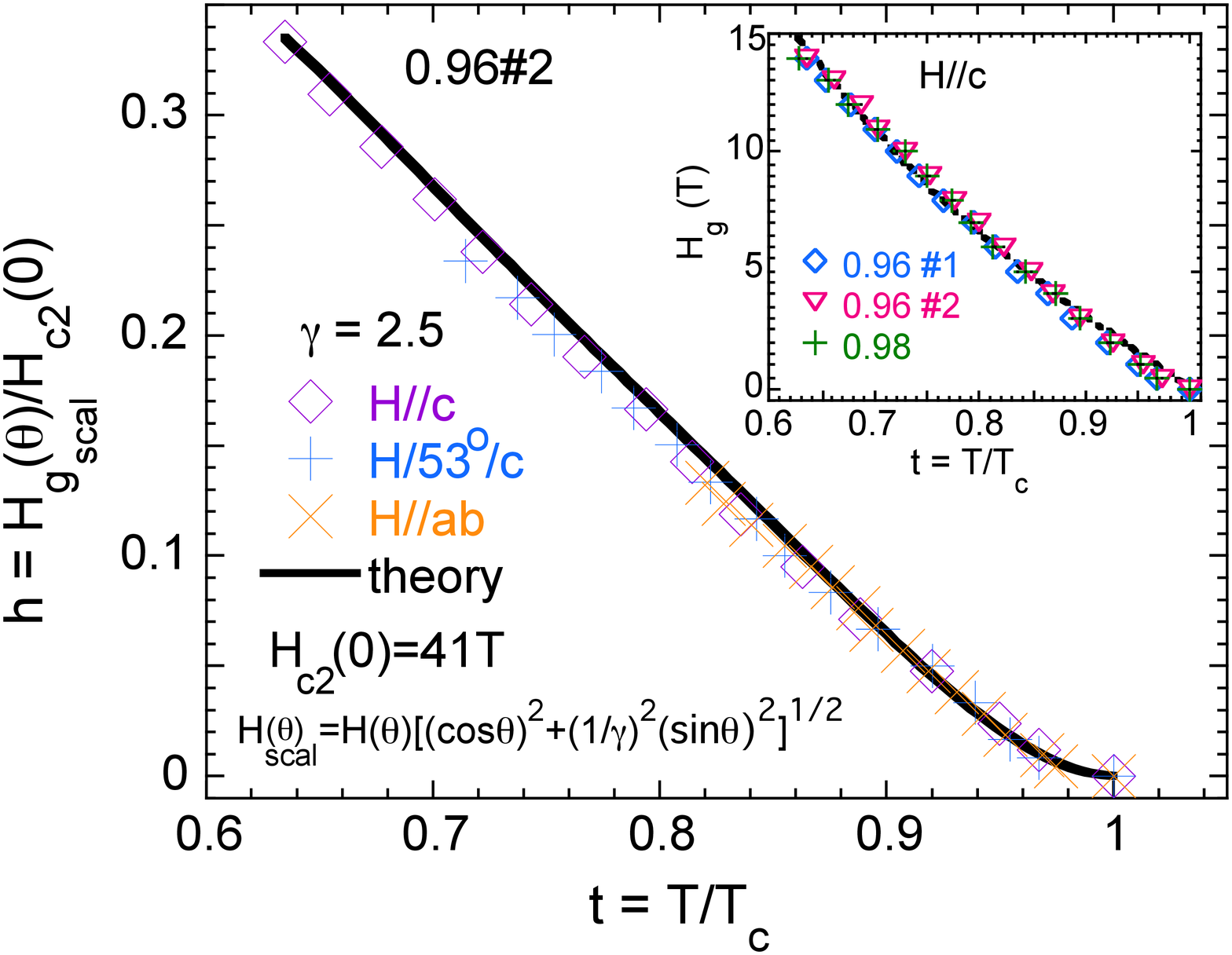}
\caption{The reduced glass field $H_g$($\theta$)/$H_{c2}(0)$ as obtained for crystal $\#$2 for the three geometries is plotted against $T$/$T_c$ after $H_g$($\theta$) is scaled accordingly the anisotropic Ginzburg-Landau theory. The solid line is a fitting to the theory. The inset show plots of $H_g$ vs. $T$/$T_c$ for the three crystals for $H$$\parallel$c-axis, where the dotted line is only a guide to the eyes.}
\label{fig5}
\end{center}
\end{figure}

Figure 5 shows a plot of the scaled reduced field, $h$=$H_g(\theta)$$_{scal}$/$H_{c2}(0)$ vs. $T$/$T_c$, where values of $H_g(\theta)$ correspond to the ($T_g$,$H$) glass line obtained for crystal $\#$2 for the three different geometries of measurements, and $H_{c2}(0)$=41T as obtained above for $H$$\parallel$c-axis. Values of $H_g(\theta)$ obtained for each direction are scaled by the angle dependent Ginzburg-Landau expression for anisotropic quantities \cite{doria}, $H_g$($\theta$)$_{scal}$ = $H_g(\theta)$[(cos($\theta$))$^2$+(1/$\gamma$)$^2$(sin($\theta$)$^2$]$^{1/2}$, where $\theta$ is the angle between $H$ and the c-axis, and $\gamma$ = 2.5, as obtained from the inset of Fig. 2b. The fact that the many different points fall in a single curve evidences that the glass line obeys the anisotropic Ginzburg-Landau theory. Since, as discussed above, we can not use the expression presented in ref.\cite{rapp1} and ref.\cite{rapp2} for the glass line to fit our data, we use an expression presented in ref.\cite{baruch2} derived for the vortex-glass line by considering the effect of disorder,
\begin{equation} 
1-t-b+2[n_p(1-t)^2 b/4\pi] [3/2-(4\pi t \sqrt{2Gi}/(n_p(1-t)^2))]=0, 
\end{equation}
where $t$=$T$/$T_c$ is the reduced temperature, $b$=$H$/$H_{c2}(0)$ is the reduced field, $n_p$ is a parameter measuring the disorder and $G_i$ is a parameter associated to the Ginzburg number measuring the strength of thermal fluctuations. The above equation was previously used to fit the irreversibility line, IL, of the crystal $\#$3 as shown in ref.\cite{jesus1}. We mention that the glass line obtained in the present study does not match the irreversibility line presented in ref.\cite{jesus1} (obtained from the zero resistivity criterion) lying little below that. The solid line in Fig.5 shows a fitting of the data (crystal $\#$2) to Eq.(4) performed by assuming $T_c$ = 19.7 K which yields the values: $n_p$ = 0.004, $G_i$$\sim$10$^{-6}$ and $H_{c2}(0)$ = 42 T (it is important to note that $H_{c2}(0)$ entering in the above equation is a fitting parameter which value is virtually equal to the value estimated above for $H{c2}(0)$ = 41 T for $H$$\parallel$c-axis which was used to obtain the values of the reduced field $h$ of the data points). Similar values of $n_p$ and $G_i$ were also observed in ref.\cite{sust2} for BaFe$_{2-x}$N$_x$As$_2$ while a smaller value of $n_p$ was obtained for NbSe$_2$ in ref.\cite{baruch2} evidencing the importance of disorder in the studied crystals. The inset of Fig.5 shows a plot of $H_g$ vs. $T$/$T_c$ (the glass line) for $H$$\parallel$c-axis as obtained for the 3 studied crystals. It is possible to see in this inset that the points corresponding to different crystals virtually fall in only one curve. Similar plots including data for the three crystals were obtained for the other two directions of $H$ with respect to the c-axis. The dotted line appearing in the inset of Fig.5 is only a guide to the eyes.	
\section{conclusions}			
In conclusion we observed that values of the activation energy obtained using the TAFF model show small deviations among the three studied crystals, while plots of the respective glass lines are virtually identical. We compared the values of $U_0$ obtained from different methods, and observed that while values of $U_0$ obtained from flux-creep data in the literature are about one order of magnitude smaller than values obtained here from the TAFF model, they are about the same order of magnitude of values of $U_b$ obtained from the modified vortex-glass model. Plots of $U_0$ as a function of temperature obtained from the modified vortex-glass model for $H$$\parallel$c-axis and $H$$\parallel$ab-planes allowed to extract the values of the glass temperature $T_g$($H$) and also of a temperature $T^*$($H$) which is of the same order of magnitude of the mean field critical temperature $T_c(H)$. An empirical expression for $U_0$ suggested by the $U_0$ vs $T$ plots within the modified vortex-glass model, allowed to obtain an universal expression for $\rho$/$\rho _n$ as a function of a scaled temperature $t_s$, which applied well to our crystal $\#$2 data. The resulting glass lines obtained for the three different geometries for each crystal are shown to obey the anisotropic Ginzburg-Landau theory, where the resulting scaled glass-line appears to be well fitted by a theory developed in ref.\cite{baruch2} for the vortex-glass line taking into account the effect of disorder. 
\ack

SSS and ADA acknowledge support from the CNPq. The work at IOP, CAS, is supported by MOST (973 project: Grants No. 2011CBA00110 and No. 2012CB821400), NSFC (11374011 and 11374346), and CAS (SPRP-B: XDB07020300)

\end{document}